\documentclass
[aps,superscriptaddress,preprint,noshowpacs,noshowkeys,nofootinbib,titlepage,twocolumn,fleqn,nobalancelastpage,tightenlines,floats]{revtex4}%
\usepackage{amssymb}
\usepackage{amsfonts}
\usepackage{amsmath}
\usepackage{graphicx}
\usepackage{float}%
\usepackage{epstopdf}
\usepackage{epsfig}
\usepackage{dcolumn}
\usepackage{bm}
\providecommand{\U}[1]{\protect\rule{.1in}{.1in}}

\begin{document}
\preprint{HEP/123-qed}
\title[Short title for running header]{Observability of the scalar Aharonov-Bohm effect inside a 3D Faraday cage with
time-varying exterior charges and masses}
\author{RY Chiao}
\affiliation{University of California, Merced, Schools of Natural Sciences and Engineering,
P.O. Box 2039, Merced, CA 95344, USA}
\author{XH Deng}
\affiliation{University of California, Merced, School of Natural Sciences, P.O. Box 2039,
Merced, CA 95344, USA}
\author{KM Sundqvist}
\affiliation{Electrical and Computer Engineering, Texas AM University, College Station, TX
77843, USA}
\author{NA Inan}
\affiliation{University of California, Merced, School of Natural Sciences, P.O. Box 2039,
Merced, CA 95344, USA}
\author{GA Munoz}
\affiliation{California State University, Fresno, CA 93740, USA}
\author{DA Singleton}
\affiliation{California State University, Fresno, CA 93740, USA}
\author{BS Kang}
\affiliation{University of California, Merced, School of Natural Sciences, P.O. Box 2039,
Merced, CA 95344, USA}
\author{LA Martinez}
\affiliation{University of California, Merced, School of Natural Sciences, P.O. Box 2039,
Merced, CA 95344, USA}
\pacs{PACS number}

\begin{abstract}
In this paper we investigate the scalar Aharonov-Bohm (AB) effect in two of
its forms, i.e., its electric form and its gravitational form. The standard
form of the electric AB effect involves having particles (such as electrons)
move in regions with zero electric field but different electric potentials.
When a particle is recombined with itself, it will have a different phase,
which can show up as a change in the way the single particle interferes with
itself when it is recombined with itself. In the case where one has
quasi-static fields and potentials, the particle will invariably encounter
fringing fields, which makes the theoretical and experimental status of the
electric AB effect much less clear than that of the magnetic (or vector) AB
effect. Here we propose using time varying fields $outside$ of a spherical
shell, and potentials $inside$ a spherical shell to experimentally test the
scalar AB effect. In our proposal a quantum system will always be in a
field--free region but subjected to a non-zero time-varying potentials.
Furthermore, our system will not be spatially split and brought back together
as in the magnetic AB experiment. Therefore there is no $spatial$ interference
and hence no shift in a $spatial$ interference pattern to observe. Rather,
there arises purely $temporal$ interference phenomena. As in the magnetic AB
experiments, these effects are non-classical. We present two versions of this
idea: (i) a Josephson temporal interferometry experiment inside a
superconducting spherical shell with a time-varying surface charge; (ii) a
two-level atom experiment in which the atomic spectrum acquires FM sidebands
when it is placed inside a spherical shell whose exterior mass is sinusoidally
varying with time. The former leads to a time-varying internal magnetic field,
and the latter leads to a time-varying gravitational redshift.

\end{abstract}
\volumeyear{year}
\volumenumber{number}
\issuenumber{number}
\eid{identifier}
\date[Date text]{date}
\received[Received text]{date}

\revised[Revised text]{date}

\accepted[Accepted text]{date}

\published[Published text]{date}

\startpage{1}
\endpage{ }
\maketitle

\section{Introduction}

In the 19th century, Faraday showed that when the exterior of a large,
enclosed cubical metallic cage (i.e., a \textquotedblleft Faraday
cage\textquotedblright) is electrified at such a high voltage that sparks
started to dart from the corners of the cage, he could still safely conduct
many sensitive electrical experiments within the cage, such as sensitive
electroscope measurements of the charge residing on the interior surface of
the cage. He found the complete absence of any charges residing on the
interior surface. Therefore in the special case of a spherical
\textquotedblleft Faraday cage\textquotedblright\ configuration, such as the
one depicted in Figure 1, one would never expect any kind of electrical
effects to be detectable inside the hollow spherical cavity which is carved
out of this metallic sphere.

But what is impossible classically is sometimes possible quantum mechanically.
For example, a 2D, cylindrical (i.e., tubular) Faraday cage was used in
Aharonov and Bohm's original paper \cite{AB effect}, in which they first
proposed the electric (or \textquotedblleft scalar\textquotedblright)
Aharonov-Bohm (AB) effect.\footnote{We do not use the term \textquotedblleft
scalar\textquotedblright\ here to refer to neutron interferometry experiments
that have been conducted in a uniform magnetic field, but reserve it to refer
to the electric and gravitational AB effects.} A metallic tube shielded an
electron passing through the tube from any exterior electric fields. However,
if a voltage pulse were to be applied to the $exterior$ of the tube only when
the electron wavepacket were to be deep in the $interior$ of the tube, then
the electron could not feel any forces during its passage through the tube.
Nevertheless, the electron would pick up the scalar AB phase shift%
\begin{equation}
\varphi(t)=\frac{e}{\hbar}\int_{0}^{t}V\left(  t^{\prime}\right)  dt^{\prime
}\label{scalar AB phase shift}%
\end{equation}
caused by the voltage\ pulse $V\left(  t\right)  $ applied to the tube whilst
the electron was deep in the interior of the Faraday cage.

The scalar (electric) AB effect is less known than the vector (magnetic)
version of the AB effect since it is harder to achieve the situation required
for the electric AB effect where fields are vanishing while the potentials are
non-zero. If one considers the fields and potentials for the cylindrical
Faraday cage used in Aharonov and Bohm's original paper, then the electron
will invariably pass through some region with a non-zero electric field,
although the field may be extremely small. To have the electron pass
\textit{only} through regions where there is no field, one needs to switch the
fields and potentials on and off completely, i.e., it is necessary to consider
the time-dependent fields and potentials described by equation
(\ref{scalar AB phase shift}), where $V(t)$ is a function with compact support.

The existence of the scalar AB effect has been questioned (see for example the
paper by Walstad \cite{Walstad}) exactly on the basis that some experimental
confirmations of the effect \cite{Oudenaarden} have the interfering electron
passing through regions where the electric field is non-zero, and thus
(potentially) one could explain the shift in an interference pattern in terms
of classical forces rather than as a quantum phase effect.

In this paper we are proposing two variants of the scalar AB effect as two
\textquotedblleft thought experiments\textquotedblright\ which address these
questions with setups where a quantum system that is influenced by potentials
is \textit{always} deep inside a field-free region of space. However, since
both of these \textquotedblleft thought experiments\textquotedblright\ involve
time-dependent potentials which have no spatial gradients, the resulting
effect on the system is not a $spatial$ interference phenomenon, as in the
vector (magnetic) AB effect, but rather a $temporal$ interference phenomenon.

From equation (\ref{scalar AB phase shift}) it can be seen that the phase in
the case of the electric AB effect involves just an open time integration as
opposed to the usual closed-path spatial line integral of the vector
(magnetic) AB effect. This opens the possibility of setting up a scalar AB
experiment where one does not split the system along different spatial paths,
as in the vector AB setup, but instead the quantum system stays at a single
location while the potential will be varying in time. Since we will not be
spatially splitting and recombining our system, no $spatial$ interference
pattern will result, and therefore no shift in the $spatial$ fringes as in the
vector (magnetic) AB effect. However, we shall see below that there can still
be a shift in the fringes of a purely $temporal$ interference pattern, or a
change of the frequency spectrum of the system.

Our proposal directly tests the scalar AB effect without any \textquotedblleft
loopholes\textquotedblright\ that would allow for the effect to be explained
any other way. The basic setup involves a spherical metallic shell (i.e., a
Faraday cage) that has an oscillating charge $Q\left(  t\right)  $ or mass
$M\left(  t\right)  $ deposited on it. Inside the shell, the potential is
spatially uniform, but is time-varying. The two systems that are placed inside
this shell are a Josephson-circuit setup and a two-level atom. In both cases,
the interior system does not move spatially and there is a complete absence of
any field (electric or gravitational), but the system will experience a
time-varying potential energy $U\left(  t\right)  $ which creates an
observable AB effect.

\section{An electric scalar AB effect via Josephson interferometry}

\begin{figure}[ptb]%
\centering
\includegraphics[height=2.025in,width=3.0592in]{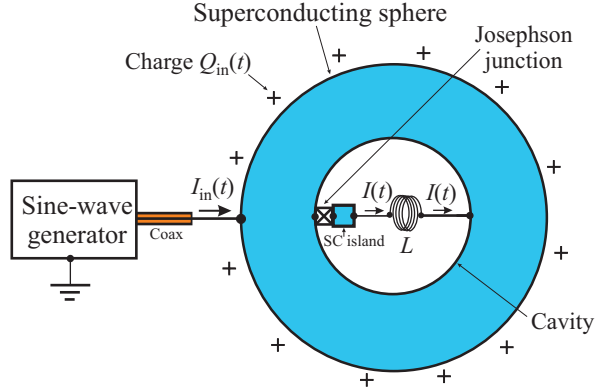}
\caption{In the spherical superconducting (SC) shell, the SC artificial atom
is formed with the hollow cavity and a superconducting wire on the horizontal
axis. A SC island is connected to the cavity with a SC wire on one end and via
a large Josephson junction on the other end. The effective inductance between
the wire and cavity can be enhanced by increasing the number of switchbacks
depicted as a solenoid. The spherical SC shell is made thick enough to prevent
magnetic flux penetration into interior.}%
\label{3DrfSQUID}%
\end{figure}

We present here a method to observe the electric scalar AB effect by a
superconducting artificial atom \cite{Clarke Review, Makhlin Review}. In order
to shield the \textquotedblleft atom\textquotedblright\ from external magnetic
fields, as shown in Figure \ref{3DrfSQUID}, it is confined within a
superconducting Faraday cage. Internal scalar potential differences arise by
sending rf-signals to the surface of the cage. Due to the shielding of the
cage, communication between the exterior and interior can only be by way of
the scalar AB phase of a Cooper pair, in the limit of electrostatics
equilibrium established instantly all over the whole Faraday cage. This
assumption satisfies the fact that the skin effect on the surface of the cage
doesn't affect the interior. It may be possible that indications of this
internal effect can also be measured by the external circuit.

\begin{figure}[ptb]%
\centering
\includegraphics[
height=1.3518in,
width=3.0805in
]%
{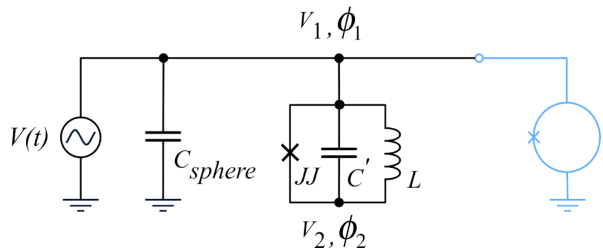}%
\caption{A simplified circuit includes three circuit loops that connected on
node 1. In the leftmost loop, a signal generator drives the system. The next
loop with the capacitor $C_{sphere}$ represents the spherical SC shell.
$C^{\prime}$ and $L$ are the effective capacitance and inductance of the SC
island, respectively. The superconducting phase $\varphi_{2}$ on the central
superconductor (node 2) and $\varphi_{1}$ on the sphere (node 1) are separated
by a Josephson junction modeled with nonlinear inductance $L_{J}$ and
effective capacitance $C_{J}$. The right loop demonstrates a simple example of
measurement, an rf-SQUID (light blue), although many possible variations of
external low-noise readout exist to detect the change of $\varphi_{1}$ in
time.}%
\label{SQUID circuit}%
\end{figure}

We demonstrate this system by considering the superconducting model circuit of
Figure \ref{SQUID circuit}. An AC voltage is introduced onto the sphere by
charging its self-capacitance. For a superconductor subject to a voltage
$V\left(  t\right)  $, a phase factor $\varphi(t)=\tfrac{2e}{\hbar}\int%
_{0}^{t}V\left(  t\right)  dt$ develops within the order parameter of the
Cooper pair condensate, $\Psi=\sqrt{\rho}e^{i\varphi}$. Furthermore, a phase
difference between the superconducting banks of the Josephson junction also
gives rise to a supercurrent according to the inverse AC Josephson effect,
i.e. the Levinsen effect \cite{Inverse ac JosephsonE}. It is therefore
possible that the scalar AB phase allows the fully enclosed superconducting
circuit to be driven by the external signal generator.

\begin{figure}[ptb]%
\centering
\includegraphics[
height=2.4353in,
width=3.039in
]%
{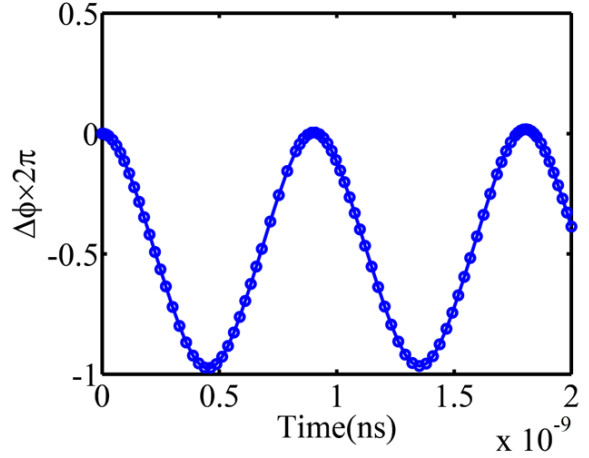}%
\caption{Phase difference across JJ $\Delta\varphi$ in terms of tme, solved
from the Lagrangian equation of motion, when $\omega_{0}\approx8.5GHz,\alpha
\approx2\times10^{-15}Wb,V_{0}=1\mu V,\omega=150MHz.$}%
\label{psi}%
\end{figure}

Neglecting possible readout schemes for our model circuit, we construct the
Lagrangian using node potentials and phases $\{V_{1},\varphi_{1};V_{2}%
,\varphi_{2}\}$ in the standard way \cite{QuantumCircuit, Devoret}. Node
phases are related to fluxes by $\varphi_{1,2}=\dfrac{2\pi}{\Phi_{0}}$
$\Phi_{1,2}=\dfrac{2\pi}{\Phi_{0}}\int\mathbf{A}_{1,2}\cdot d\mathbf{x,}$
where flux quanta are $\Phi_{0}=h/2e$. Omitting bias conditions, we find
\begin{align}
\mathcal{L}\left(  \varphi_{2},t\right)   &  =\frac{C_{sphere}}{2}V_{1}%
^{2}\nonumber\\
&  \hspace{0.2in}+\frac{C_{\Sigma}}{2}\left(  V_{1}-V_{2}\right)
^{2}\nonumber\\
&  \hspace{0.2in}-\frac{1}{2L}\left(  \Phi_{1}-\Phi_{2}\right)  ^{2}%
\nonumber\\
&  \hspace{0.2in}+E_{J}\cos\alpha\left(  \Phi_{1}-\Phi_{2}\right) \\
& \nonumber\\
&  =\left(  \frac{\hbar}{2e}\right)  ^{2}\left[  \tfrac{1}{2}C_{sphere}%
\dot{\varphi}_{1}^{2}\right. \nonumber\\
&  \hspace{0.2in}+\frac{C_{\Sigma}}{2}\left(  \dot{\varphi}_{2}-\dot{\varphi
}_{1}\right)  ^{2}\nonumber\\
&  \hspace{0.2in}-\frac{1}{2L}\left(  \varphi_{2}-\varphi_{1}\right)
^{2}\nonumber\\
&  \hspace{0.2in}\left.  +E_{J}\cos\left(  \varphi_{2}-\varphi_{1}\right)
\right]
\end{align}
Considering $V_{1}$ is driven $\left(  V_{1}=V_{0}\sin\omega t\right)  $, the
interior phase $\Delta\varphi=\varphi_{2}-\varphi_{1}$ and $\Delta\dot
{\varphi}=\dot{\varphi}_{2}-\dot{\varphi}_{1}$ become the independent
variables in the Lagrangian. This allows us to get the following equation of
motion with driving on $\dot{\varphi}_{1}$.%
\begin{align}
0 &  =\Delta\ddot{\varphi}+\omega_{C}^{2}\Delta\varphi+\left(  \frac{2e}%
{\hbar}\right)  ^{2}\frac{E_{J}}{C_{\Sigma}}\sin\Delta\varphi\nonumber\\
&  \hspace{0.2in}+\left(  \frac{2e}{\hbar}\right)  ^{2}\frac{C_{g}V_{0}\omega
}{C_{\Sigma}}\cos\omega t\label{equation of motion}%
\end{align}
where $\Delta\varphi=\left(  \varphi_{2}-\varphi_{1}\right)  $, $\omega
_{C}=\dfrac{1}{LC^{\prime}}$. Here $C_{g}$ is the effective gate capacitance
of the SC island and $C_{\Sigma}$ is the total capacitance of the system. We
adopt the following initial conditions, which assume an initial steady state
throughout the system.
\begin{equation}
\left\{
\begin{array}
[c]{c}%
\varphi_{2}\left(  t=0\right)  =\varphi_{1}\left(  t=0\right) \\
\dot{\varphi}_{2}\left(  t=0\right)  =\dot{\varphi}_{1}\left(  t=0\right)
\end{array}
\right.
\end{equation}
Figure \ref{psi} depicts the calculated dynamics of $\Delta\varphi$. When the
floating part node 2 is initially grounded and the internal circuit is cooled
down to ground state, a signal of $V_{1}$ drives $\Delta\varphi$ starting at
some moment according to the equation of motion (\ref{equation of motion}).
Hence an interior supercurrent is being driven after the signal is turned off
at some point and causes a nonvanishing $\Delta\varphi$ because of the
Josephson effect, which reflects on changing $\varphi_{1}$ from the internal
circuit. An rf-SQUID (light blue in Figure \ref{SQUID circuit}) at the read
out port picks up the time dependent phase $\varphi_{1}$ and converts it into
time dependent magnetic flux which can be detected from an exterior circuit.
We have therefore demonstrated that the external generator affects the fully
enclosed superconducting circuit.

\begin{figure}[pth]%
\centering
\includegraphics[
height=2.3142in,
width=3.1652in
]%
{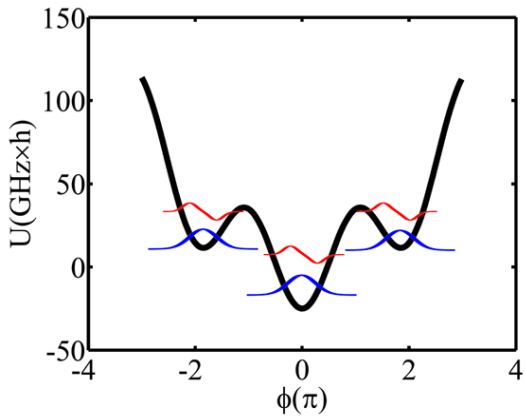}%
\caption{A plot of potential energy for $E_{L}=1GHz\times h,E_{J}=25GHz\times
h.$ Blue is the ground state for each well while red is the first excited
state. Inter-well tunneling may occur, in which microwave photons may be
emitted or absorbed. }%
\label{Potential Energy}%
\end{figure}

Also notice that the internal Josephson junction circuit, with the possible
addition of current or flux biasing, shares a similar circuit topology to
phase or flux qubits, respectively \cite{Phase qubit, Flux qubit}. In any
case, when $E_{J}\gg E_{C}=\tfrac{\left(  2e\right)  ^{2}}{2C_{\Sigma}}$,
phase is a good quantum number. The anharmonic potential energy, $U\left(
{\Delta\varphi}\right)  =\left(  {\tfrac{\hbar}{{2e}}}\right)  ^{2}\tfrac
{1}{{2L}}\left(  {\Delta\varphi}\right)  ^{2}-E_{J}\cos\left(  {\Delta\varphi
}\right)  $ is depicted as in Figure \ref{Potential Energy} and gives
quantized energy levels similar to an atom occupying the local minimas of this
periodic potential. This artificial atomic system, as we have just determined,
is affected by the external modulating voltage. Note that the electric
potential becomes $\partial/\partial\varphi$ which is the conjugate variable
of ${\Delta\varphi}$ in the Hamiltonian. Driving on the potential could
displace the phase from ${\Delta\varphi=0}$ to ${\Delta\varphi=2\pi}$. A
$2\pi$ phase in the internal circuit corresponds to a flux quantum in the
cavity generated by the persistent current going through the central SC wire.
After releasing the driving, the system may stay in the ${\Delta\varphi=2\pi}$
state until there is a relaxation back to ${\Delta\varphi=0}$, which can be
detected via measurement of the rf-SQUID (rightmost loop in Figure
\ref{SQUID circuit}). If the driving finishes one cycle from ${\Delta
\varphi=0}$ to ${\Delta\varphi=0}$, the magnetic flux quantum comes back to
\textit{zero} in the cavity. Therefore, the internal magnetic flux becomes the
consequence of a temporal scalar AB effect.%

Here we make the following observations:

(1) The supercurrent is created due interference between two paths for the
propagation of the SC phase to node-2. One of the paths is via the JJ and the
other is via the solenoid.

(2) The scalar AB phase discussed here is only for a single Cooper pair
instead of a bulk system. As pointed out earlier, the phase difference drives
the supercurrent that in principle can be detected. Considering the bulk
system, the AB phase factor should include the phase from all Cooper pairs,
$\varphi_{C.P.}\left(  t\right)  =\frac{2e}{\hbar}\int_{0}^{t}N_{C.P.}\left(
t\right)  V\left(  t\right)  dt$, and electrons,$\ \varphi_{els}\left(
t\right)  =\frac{e}{\hbar}\int_{0}^{t}N_{els}\left(  t\right)  V\left(
t\right)  dt$. Here $N_{C.P.}\left(  t\right)  $ is the total number of Cooper
pairs (which is not conserved in principle) and $N_{els}\left(  t\right)  $ is
the total number of electrons of the whole system. Also, AB phase factor from
ionic lattice should be included as well $\varphi_{ion}\left(  t\right)
=\frac{-e}{\hbar}\int_{0}^{t}N_{ion}\left(  t\right)  V\left(  t\right)  dt$.
For the case of a 3D rf-SQUID confined within the Faraday cage (see Figure
\ref{3DrfSQUID}), there will arise time-varying charge imbalances between the
charge of the fixed ionic lattice and the charge of the mobile Cooper pairs on
the SC island, which will give rise to a nonzero total AB phase developed in
the internal circuit. The AB effect here merely drives the Cooper pair
condensate and generates a magnetic field in the internal cavity. This is the
physical reaction caused by the scalar AB phase that we expect here by
theoretical analysis.

(3) Driving the SC atom away from ${\Delta\varphi=0}$ to ${\Delta\varphi=2}%
\pi$ becomes analogous to the case of the ionization of an atom. Consequently,
we can expect that treating this perturbation results in a nonlocal
photoelectric effect.

\section{A gravitational AB phase shift observable as a time-dependent
gravitational redshift}

Here we consider the problem of a two-level atom that is undergoing a
time-dependent gravitational redshift when the atom is placed inside a
time-varying spherical mass shell $M(t)$ (see Figure 5) Again, as in the
electric case, the gravitational scalar potential will be uniform everywhere
within the interior of a mass shell, so that no gravitational force will be
experienced by the atom. Nevertheless, there can in general arise a scalar AB
phase \cite{Hohensee}, which arises from the Newtonian gravitational scalar
potential $\Phi$%
\begin{equation}
\varphi\left(  t\right)  =\frac{1}{\hbar}\int_{0}^{t}m\left(  t^{\prime
}\right)  \Phi\left(  t^{\prime}\right)  dt^{\prime}\label{GR AB phase}%
\end{equation}
where $m\left(  t^{\prime}\right)  $ is the time-varying rest mass of the
quantum system that is acquiring this phase shift. The Newtonian gravitational
scalar potential for a time-varying mass shell $M\left(  t\right)  $, such as
that associated with the sinusoidally time-varying charge $Q\left(  t\right)
$ on the surface of the shell depicted in Figure 1, is given by%
\begin{align}
\Phi\left(  t\right)   &  =-\frac{GM\left(  t\right)  }{r_{0}}\nonumber\\
&  =-\frac{G\left(  M_{0}+M_{1}\cos\omega t\right)  }{r_{0}}%
\label{DC and AC components of M(t)}%
\end{align}
where $G$ is Newton's constant, $r_{0}$ is the radius of the mass shell,
$M_{0}$ is the DC component of the mass shell, and $M_{1}$ is the amplitude of
the AC component of the time-varying mass shell. It goes without saying that
the mass-to-charge ratio of the electron is so tiny that the AC component
associated with $M_{1}$ will be extremely small, so that there would be no
hope for any practical laboratory experiment in connection with the Faraday
cage configuration shown in Figure 5. However, we are concerned here with the
problem of whether \emph{in principle} the gravitational AB phase exists or
not, so that a \textquotedblleft thought experiment\textquotedblright\ would
suffice here.

Nevertheless, there exist astrophysical situations, such as in the case of an
exploding mass shell of a supernova, where the radius $r_{0}\left(  t\right)
$ is the time-dependent quantity rather than the mass $M\left(  t\right)  $.
Then the gravitational AB phase shift may be large enough to be seen in
practice. In any case, although the gravitational scalar potential may vary
with time, nevertheless it must be independent of the position of the field
point within the interior of the mass shell. This follows from Gauss's
theorem. Therefore the atom within the mass shell experiences no classical
forces. However, it can experience a nonzero quantum phase shift arising from
the gravitational AB effect.%

\begin{figure}[th]%
\centering
\includegraphics[height=2.3in,width=3.2in]{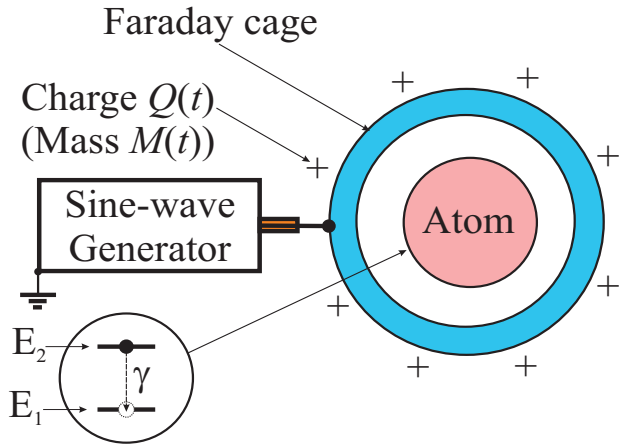}%
\caption{Two-level atom inside a spherical mass shell with a time-dependent
mass $M(t)$ that arises from a time-dependent charge $Q(t)$. The mass $M(t)$
results a time-dependent gravitational redshift that leads to observable FM
sidebands in the spectrum of the atom.}%
\end{figure}

The rest mass of the excited state of an atom or of a nucleus will be larger
than the rest mass of the ground state of this atom or nucleus. This follows
from Einstein's equation%
\begin{equation}
E=mc^{2}\label{E=mc^2}%
\end{equation}
In quantum mechanics, Einstein's equation becomes%
\begin{equation}
mc^{2}=\left\langle \psi\left(  t\right)  \right\vert i\hbar\frac{\partial
}{\partial t}\left\vert \psi\left(  t\right)  \right\rangle
\label{rest mass is expectation value of energy}%
\end{equation}
Thus the rest mass $m$ in relativity is the expectation value of the energy
operator $i\hbar\partial/\partial t$ in quantum mechanics. For example, when
an atom is in a stationary state $\left\vert \psi\left(  t\right)
\right\rangle =\left\vert \Psi_{E}\right\rangle \exp\left(  -iEt/\hbar\right)
$, it follows from (\ref{rest mass is expectation value of energy}) that%
\begin{equation}
mc^{2}=E\left\langle \Psi_{E}\right.  \left\vert \Psi_{E}\right\rangle
=E\label{quantum E= mc^2}%
\end{equation}
and thus we recover Einstein's equation (\ref{E=mc^2}). We shall assume that
(\ref{rest mass is expectation value of energy}) holds in general for open
quantum systems, such as that of an atom inside a time-varying mass shell
depicted in Figure 5.

Now from the expression for the gravitational AB phase (\ref{GR AB phase}), we
expect that the atom in a stationary state with an energy $E$ inside the mass
shell will pick up an AB phase factor, so that%
\begin{equation}
\left\vert \psi\left(  t\right)  \right\rangle =\left\vert \Psi_{E}%
\right\rangle \exp\left(  -iEt/\hbar\right)  \exp\left(  -i\varphi\left(
t\right)  \right)
\end{equation}
Substituting this into (\ref{rest mass is expectation value of energy}), we
find%
\begin{equation}
mc^{2}=\left\langle \Psi_{E}\right\vert \{E+\hbar\dot{\varphi}\}\left\vert
\Psi_{E}\right\rangle =E+\hbar\dot{\varphi}%
\label{generalized Einstein relation}%
\end{equation}
which is an application of (\ref{rest mass is expectation value of energy}) to
the case of a time-dependent environment, such as the time-varying mass shell.
To find $\dot{\varphi}$, let us take the time derivative of the expression for
the AB phase (\ref{GR AB phase}). Then one obtains the relationship%
\begin{equation}
\dot{\varphi}=\frac{1}{\hbar}m\left(  t\right)  \Phi\left(  t\right)
\label{instantaneous frequency of wavefunction}%
\end{equation}
The physical meaning of this relationship is that the \textquotedblleft
instantaneous\textquotedblright\ frequency $\dot{\varphi}$ associated with the
modulation of the phase of the atomic wavefunction due to an \textquotedblleft
instantaneous\textquotedblright\ change in the gravitational potential energy
$m\left(  t\right)  \Phi\left(  t\right)  $ of the atom inside the mass shell,
leads to an "instantaneous" energy change of the energy level of the atom
given by%
\begin{equation}
\delta E=\hbar\dot{\varphi}=m\left(  t\right)  \Phi\left(  t\right)
\end{equation}
Upon substitution of $\dot{\varphi}$ from
(\ref{instantaneous frequency of wavefunction}) back into
(\ref{generalized Einstein relation}), it follows that%
\begin{equation}
mc^{2}=E+m\left(  t\right)  \Phi\left(  t\right)
\end{equation}
from which we infer that%
\begin{equation}
E=mc^{2}\left(  1-\frac{\Phi}{c^{2}}\right) \label{E= mc^2 (1-Phi/c^2)}%
\end{equation}
which leads to the conclusion that the rest mass of quantum mechanical systems
may be increased due to its external gravitational environment.

If we now take the difference in the upper and lower energy levels of the
two-level atom and set it equal to the frequency of an emitted photon $f_{0}$
times Planck's constant $h$, we will find that%
\begin{equation}
hf_{0}=hf\left(  1-\frac{\Phi}{c^{2}}\right)
\end{equation}
Since $\Phi$ is a negative quantity, this implies that the photon of energy
$hf$ as seen by an observer at infinity will be smaller in energy than the
photon of energy $hf_{0}$ as seen by an observer near to the atom. Thus we
have recovered Einstein's gravitational redshift starting from the
gravitational AB phase shift.

Now if the potential $\Phi$ were to be time varying due to changes in the mass
shell, the gravitational red shift would be changed by the time variation of
$\Phi$. If the time variation were to be sinusoidal, then we would expect the
emission and absorption spectrum of the atom to undergo FM modulation. To see
this, let us assume that the states of a two-level atom are represented by
$\left\vert i\right\rangle $ for the initial state and by $\left\vert
f\right\rangle $ for the final state. Then Fermi's Golden Rule states that the
rate of transitions $w_{f\leftarrow i}$\ between these two states will be
given by the absolute square of the transition matrix element connecting the
initial and final states, i.e.,%
\begin{equation}
w_{f\leftarrow i}\propto\left\vert \left\langle f\right\vert H^{\prime
}\left\vert i\right\rangle \right\vert ^{2}\label{Fermi golden rule}%
\end{equation}
where $H^{\prime}$ is the time-dependent perturbation that causes the
transitions to occur. (For the present purposes, we ignore the proportionality
constant and the density of final states in Fermi's Golden Rule.)

Let us first consider the effect of the $electric$ scalar AB effect on the
transitions between the initial and final states of a $charged$ atom, i.e., an
ion, within the spherical shell of the Faraday cage. The electric AB phase is
given by%
\begin{equation}
\varphi_{q}\left(  t\right)  =\frac{q}{\hbar}\int_{0}^{t}V\left(  t^{\prime
}\right)  dt^{\prime}%
\end{equation}
where $q$ is the charge of the ion. Since the initial and final states of the
transition must have the same charge $q$ (which follows from Wigner's charge
superselection rule), it follows that the electric AB phase factors in the
transition matrix elements in Fermi's Golden Rule must cancel out, i.e.,%
\begin{align}
w_{f\leftarrow i} &  \propto\left\vert \left\langle f\right\vert
e^{+i\varphi_{q}(t)}H^{\prime}e^{-i\varphi_{q}(t)}\left\vert i\right\rangle
\right\vert ^{2}\nonumber\\
&  =\left\vert \left\langle f\right\vert H^{\prime}\left\vert i\right\rangle
\right\vert ^{2}\label{invisibility of electric AB effect in atoms}%
\end{align}
since $[H^{\prime},V\left(  t\right)  ]=0$. From
(\ref{invisibility of electric AB effect in atoms}), we conclude that the
electric AB effect cannot be observed in the spectroscopy of any charged
atomic system.

However, this is $not$ true for the gravitational AB effect. This is because
of the fact that the $rest$ $mass$ of an excited atom will be greater than the
$rest$ $mass$ of the unexcited atom. (Recall that there exists no
superselection rule for mass, unlike for the case of charge.)

Now from Einstein's equation (\ref{E=mc^2}),\ it follows that there exists a
rest mass difference between the final and the initial states of the two-level
atom, which is given by%
\begin{equation}
\Delta m=\frac{\Delta E}{c^{2}}=\frac{E_{f}-E_{i}}{c^{2}}%
\label{difference in rest masses}%
\end{equation}
where $E_{f}$ is the energy level of the final state and $E_{i}$ is the energy
level of the initial state. Therefore from (\ref{GR AB phase}), we see that
the difference in the gravitational AB phase picked up by the final state and
the phase picked up the initial state will no longer vanish, but will differ
by the amount%
\begin{align}
\Delta\varphi\left(  t\right)   &  =\frac{\Delta m}{\hbar}\int_{0}^{t}%
\Phi\left(  t^{\prime}\right)  dt^{\prime}\nonumber\\
&  =\frac{E_{f}-E_{i}}{\hbar c^{2}}\int_{0}^{t}\Phi\left(  t^{\prime}\right)
dt^{\prime}\label{difference in AB phases}%
\end{align}
where, to a a first approximation, the energy difference $E_{f}-E_{i}$ is
independent of $\Phi$.

As a simple example of how the difference in the gravitational AB phases in
the initial and final states can lead to an observable AB interference effect,
let us consider a superposition of the initial and final states which is
initially given by%
\begin{equation}
\left\vert \Psi\left(  t=0\right)  \right\rangle \propto\left\vert
i\right\rangle +\left\vert f\right\rangle
\end{equation}
After a time $t=T$, this superposition will evolve to pick up phase factors,
viz.,%
\begin{align}
\left\vert \Psi\left(  t=T\right)  \right\rangle  &  \propto\left\vert
i\right\rangle e^{-\frac{i}{\hbar}E_{i}T}e^{-i\varphi_{i}T}\nonumber\\
&  +\left\vert f\right\rangle e^{-\frac{i}{\hbar}E_{f}T}e^{-i\varphi_{f}%
T}\label{superposition state at t=T}%
\end{align}
where%
\begin{equation}
\varphi_{i}\left(  T\right)  =\frac{E_{i}}{\hbar c^{2}}\int_{0}^{T}\Phi\left(
t^{\prime}\right)  dt^{\prime}%
\end{equation}%
\begin{equation}
\varphi_{f}\left(  T\right)  =\frac{E_{f}}{\hbar c^{2}}\int_{0}^{T}\Phi\left(
t^{\prime}\right)  dt^{\prime}%
\end{equation}
are the gravitational AB phases picked up by the initial and final states,
respectively. From (\ref{difference in AB phases}) and
(\ref{superposition state at t=T}) it follows that%
\begin{equation}
\left\vert \Psi\left(  t=T\right)  \right\rangle \propto\left\vert
i\right\rangle +\left\vert f\right\rangle e^{-\frac{i}{\hbar}\left(
E_{f}-E_{i}\right)  T}e^{-i\Delta\varphi T}%
\end{equation}
where $\Delta\varphi\left(  T\right)  =\varphi_{f}\left(  T\right)
-\varphi_{i}\left(  T\right)  $ is the difference of the two AB phases. From
(\ref{GR AB phase}) and (\ref{DC and AC components of M(t)}), we see that%
\begin{equation}
\varphi_{i}\left(  T\right)  =\alpha_{i}\sin\omega
t\label{phase modulation of initial state}%
\end{equation}%
\begin{equation}
\varphi_{f}\left(  T\right)  =\alpha_{f}\sin\omega
t\label{phase modulation of final state}%
\end{equation}
where $\alpha_{i}=GM_{1}m_{i}/\hbar\omega r_{0}$ and $\alpha_{f}=GM_{1}%
m_{f}/\hbar\omega r_{0}$ are the FM modulation parameters for the initial and
final wavefunctions, respectively, of the two-level atom. The sinusoidal
modulations of the phases given by (\ref{phase modulation of initial state})
and (\ref{phase modulation of final state}) will lead to many FM harmonics of
the frequency $\omega$ via the Jacobi-Anger expansion of the wavefunctions of
the two-level atom (see Appendix A). For large values of $\alpha_{i}$ and
$\alpha_{f}$, the dominant upper and lower FM sidebands of the FM-modulated
wavefunctions of the atom will occur at the frequency shifts $\pm\alpha
_{i}\omega$ and $\pm\alpha_{f}\omega$ away from their usual frequencies of
$E_{i}/\hbar$ and $E_{f}/\hbar$.

It follows that the usual energy-conservation-enforcing delta function in the
Fermi Golden rule will be modified from the usual two-level atom resonance
condition not only by the usual gravitational red shift stemming from the DC
component of the mass shell $M_{0}$, but it will also be modified due to the
FM sidebands that arise from the AC component of the time-varying mass shell
$M\left(  t\right)  $. The bottom line of this analysis is that the usual
absorption or emission line of the two-level atom will be split into upper and
lower FM sideband frequencies occurring on either side of the unsplit line of
the atom with frequency shifts of $\pm(\Delta\alpha)\omega$, where%
\begin{align}
\Delta\alpha &  =\left(  \alpha_{f}-\alpha_{i}\right) \nonumber\\
&  =\frac{GM_{1}\left(  m_{f}-m_{i}\right)  }{\hbar\omega r_{0}}%
\end{align}
is the difference between the FM modulation parameters of the final state and
the initial state that stems from the difference in their rest masses,
$m_{f}-m_{i}$.

\section{Conclusions}

We conclude from the above two \textquotedblleft
counter-examples\textquotedblright\ that the claims of the non-existence of
the scalar AB effect are false. Although these two \textquotedblleft
counter-examples\textquotedblright\ are by nature merely \textquotedblleft
thought experiments,\textquotedblright\ they do establish the $existence$ of
the electric AB effect and the $existence$ of the gravitational AB effect
\emph{in principle}. However, they may ultimately lead to actual experiments
in the laboratory in the Josephson interferometry case, and to actual
observational evidence in astrophysical settings in the gravitational redshift case.

Finally, we note that the magnetic (vector) AB effect as observed by Tonomura
\cite{Tonomura} using ferromagnetic toroids in electron interference
experiments, are obviously topological in nature. However, the electric and
gravitational (scalar) AB effects that are predicted to occur here inside the
metallic shells of Figures 1 and 5, are obviously non-topological in nature.

\section{Appendix A: Bloch's theorem for scalar potentials that are periodic
in time}

Consider the general case in which the potential energy inside the Faraday
cages depicted in Figures 1 and 5 satisfy the periodicity condition in time%
\begin{equation}
U\left(  t+T\right)  =U\left(  t\right)
\label{temporally periodic potential energy}%
\end{equation}
where the period $T=2\pi/\omega$ is that of an arbitrary periodic waveform
generator that replaces the sine-wave generators in Figures 1 and 5. Then it
is apparent that this temporal periodicity condition is mathematically
identical to the spatial periodicity condition%
\begin{equation}
U\left(  x+a\right)  =U\left(  x\right)
\label{spatially periodic potential energy}%
\end{equation}
that applies to a 1D crystalline lattice with a lattice constant $a$.

Bloch's theorem \cite{Kittel} then tells us that the wavefunction inside the
1D crystalline lattice is given by%
\begin{equation}
\psi\left(  x\right)  =e^{ipx/\hbar}u_{p}\left(  x\right)
\label{Bloch's theorem spatial form}%
\end{equation}
where $p$ is the \textquotedblleft crystal momentum\textquotedblright\ or
\textquotedblleft quasi-momentum\textquotedblright, and where%
\begin{equation}
u_{p}\left(  x+a\right)  =u_{p}\left(  x\right)
\end{equation}
is a periodic function of $x$ within the spatial crystalline lattice.

Similarly, the temporal version of Bloch's theorem (also known as
\textquotedblleft Floquet's theorem\textquotedblright) is given by%
\begin{equation}
\psi\left(  t\right)  =e^{-iEt/\hbar}u_{E}\left(  t\right)
\label{Bloch's theorem temporal form}%
\end{equation}
where $E$ is the \textquotedblleft crystal energy\textquotedblright\ or
\textquotedblleft quasi-energy\textquotedblright\ \cite{Zeldovich}%
\cite{Silveri}, and where%
\begin{equation}
u_{E}\left(  t+T\right)  =u_{E}\left(  t\right)
\end{equation}
is a periodic function of $t$ within a certain \textquotedblleft temporal
crystalline lattice.\textquotedblright\ \cite{Wilczek}\ (We have suppressed
the spatial dependence of the wavefunction $\psi\left(  t\right)  $ and of the
periodic function $u_{E}\left(  t\right)  $ as being understood in
(\ref{Bloch's theorem temporal form}). This also applies to all of the
following expressions for $\psi\left(  t\right)  $ and $u_{E}\left(  t\right)
$). Both the \textquotedblleft crystal momentum\textquotedblright\ and the
\textquotedblleft crystal energy\textquotedblright\ are \emph{physically
observable quantities} that obey \emph{conservation laws}, because of the
discrete translational symmetry of the crystalline systems in $x$ and in $t$,
respectively, which follow from the translational symmetry of
(\ref{temporally periodic potential energy}) and
(\ref{spatially periodic potential energy}).

Since $u_{E}\left(  t\right)  $ is a periodic function of time with a period
$T $, it can be expanded by Fourier's theorem into a Fourier series expansion%
\begin{equation}
u_{E}\left(  t\right)  =\sum_{n=-\infty}^{+\infty}c_{n}\exp\left(  -in\omega
t\right) \label{Fourier series expansion of u(t)}%
\end{equation}
where $c_{n}$ are the Fourier coefficients of $u_{E}\left(  t\right)  $, and
where $\omega=2\pi/T$ is the frequency of periodic charge waveform $Q\left(
t\right)  $ that is being injected onto the surface of the Faraday cage in
Figures 1 and 5 by the arbitrary periodic waveform generator. Substituting
(\ref{Fourier series expansion of u(t)}) into
(\ref{Bloch's theorem temporal form}), one finds that%
\begin{align}
\psi\left(  t\right)   &  =\sum_{n=-\infty}^{+\infty}c_{n}\exp\left(
-i\left(  E+n\hbar\omega\right)  t/\hbar\right) \nonumber\\
&  =\sum_{n=-\infty}^{+\infty}c_{n}\exp\left(  -iE_{n}t/\hbar\right)
\label{psi expanded in terms of quasi-energies}%
\end{align}
so that we conclude that%
\begin{equation}
E_{n}=E+n\hbar\omega\label{quasi-energy level spectrum indexed by n}%
\end{equation}
which describes the \textquotedblleft quasi-energy\textquotedblright\ levels
\cite{Zeldovich}\ of any charged quantum system inside the cavity of a Faraday
cage which is being driven by an arbitrary periodic waveform.

Note that this derivation of the spectrum of quasi-energy levels
(\ref{quasi-energy level spectrum indexed by n}) applies to the case of $any$
periodic potential energy function $U\left(  t\right)  $ . However, let us now
consider the important special case of a $sinusoidal$ time variation of
$U\left(  t\right)  $.

The wavefunction of a quantum system inside the Faraday cage such as the ones
depicted in Figures 1 and 5, will be phase modulated by the time-varying
potential energy in accordance with the time-dependent Schr\"{o}dinger
equation%
\begin{equation}
i\hbar\frac{\partial\psi}{\partial t}=H\psi=(H_{0}+U(t))\psi
\end{equation}
where the $H$ is the total Hamiltonian, $H_{0}$ is the unperturbed
Hamiltonian, and $U\left(  t\right)  $ is the potential energy of the quantum
system inside the spherical shell, which results, for example, from the
injection of the charge $Q\left(  t\right)  $ onto the surface of the
spherical metallic shell (i.e., Faraday cage) in Figure 1. Note that $U\left(
t\right)  $ will be independent of the position of any field point in the
volume within the shell. Thus%
\begin{equation}
U\left(  t\right)  =U_{0}\cos\omega t\label{U(t)}%
\end{equation}
for the case of an oscillating charge $Q\left(  t\right)  =Q_{0}\cos\omega t$
which is $exterior$ to the Faraday cage. However, any field arising from the
spatial gradients of $U\left(  t\right)  $ $interior$ to the space containing
the quantum system is zero.

Now we shall assume that the quantum system is initially in an unperturbed
eigenstate of the unperturbed Hamiltonian $H_{0}$, i.e.,%
\begin{equation}
H_{0}\psi=E\psi
\end{equation}
where $E$ is the unperturbed energy level of the system. Since $\left[
H_{0},U(t)\right]  =0$, it follows that the solution to the time-dependent
Schr\"{o}dinger equation is%
\begin{align}
\psi\left(  t\right)   &  =\psi\left(  0\right)  e^{-\frac{i}{\hbar}\left(
Et+\int_{0}^{t}U(t^{\prime})dt^{\prime}\right)  }\nonumber\\
&  =\psi\left(  0\right)  e^{-\frac{i}{\hbar}Et}e^{-i\varphi\left(  t\right) }%
\end{align}
where $\varphi(t)$ is the phase shift of the wavefunction of the system which
is associated with the scalar AB effect, i.e.,%
\begin{equation}
\varphi\left(  t\right)  =\frac{1}{\hbar}\int_{0}^{t}U\left(  t^{\prime
}\right)  dt^{\prime}\label{electron AB phase shift}%
\end{equation}
in agreement with (\ref{scalar AB phase shift}) and (\ref{GR AB phase}). Using
the explicit functional form of (\ref{U(t)}) in order to evaluate this
integral, we find that%
\begin{equation}
\varphi\left(  t\right)  =\alpha\sin\omega t
\end{equation}
where the \textquotedblleft FM depth of modulation\textquotedblright\ $\alpha$
parameter is defined as follows:%
\begin{equation}
\alpha=\frac{U_{0}}{\hbar\omega}%
\end{equation}

Thus we find that the wavefunction of the system in the presence of the
$interior$ potential energy $U\left(  t\right)  $, which is\thinspace caused,
for example, by the $exterior$ charge $Q\left(  t\right)  $, will have the
form%
\begin{equation}
\psi\left(  t\right)  =\psi\left(  0\right)  e^{-\frac{i}{\hbar}Et}%
e^{-i\alpha\sin\omega t}\label{FM modulated wf}%
\end{equation}
Now from the generating function for Bessel functions, one obtains the
Jacobi-Anger expansion \cite{Abramowitz}%
\begin{equation}
e^{-i\alpha\sin\omega t}=\sum_{n=-\infty}^{\infty}J_{n}\left(  \alpha\right)
e^{-in\omega t}\label{exponential form of Bessel identity}%
\end{equation}
where $J_{n}\left(  \alpha\right)  $ is the $n^{th}$ order Bessel function of
the argument $\alpha$. The meaning of the index $n$ is that it denotes the
$n^{th}$ harmonic sideband of the phase modulated wavefunction, which will end
up modifying the quasi-energy level structure of the quantum system. Positive
values of $n$ will correspond to upshifted-frequency sidebands, and negative
values of $n$ to downshifted-frequency sidebands in the quasi-energy spectrum.

Substituting the Jacobi-Anger expansion into the wavefunction
(\ref{FM modulated wf}), we conclude that%
\begin{align}
\psi\left(  t\right)   &  =\psi\left(  0\right)  e^{-\frac{i}{\hbar}Et}%
\sum_{n=-\infty}^{\infty}J_{n}\left(  \alpha\right)  e^{-in\omega
t}\nonumber\\
&  =\psi\left(  0\right)  \sum_{n=-\infty}^{\infty}J_{n}\left(  \alpha\right)
e^{-\frac{i}{\hbar}\left(  E+n\hbar\omega\right)  t}\nonumber\\
&  =\psi\left(  0\right)  \sum_{n=-\infty}^{\infty}J_{n}\left(  \alpha\right)
e^{-\frac{i}{\hbar}E_{n}t}%
\end{align}
where the quasi-energy levels $E_{n}$ are once again given by the expression%
\begin{align}
E_{n}  & =E+n\hbar\omega\label{energy level of nth FM sideband}\\
& \nonumber
\end{align}

\section{Acknowlegements}

We thank Profs. Dan Stamper-Kurn and Chih-Chun Chien for helpful discussions.


\begin{thebibliography}{99}                                                                                               %
\bibitem {AB effect}Y. Aharonov and D. Bohm, \textquotedblleft Significance of
electromagnetic potentials in the quantum theory,\textquotedblright\ Phys.
Rev. \textbf{115}, 485 (1959).

\bibitem {Walstad}A. Walstad, \textquotedblleft A Critical Reexamination of
the Electrostatic Aharonov-Bohm Effect,\textquotedblright\ Int. J. Theor.
Phys. \textbf{49}, 2929-2934 (2011).

\bibitem {Oudenaarden}Van Oudenaarden, A., Devoret, M.H., Nazarov, Yu.V.,
Mooij, J.E.: Nature \textbf{391}, 768--770 (1998).

\bibitem {Clarke Review}John Clarke \& Frank K. Wilhelm, \textquotedblleft
Superconducting quantum bits\textquotedblright, Nature 453, 1031(2008).

\bibitem {Makhlin Review}Yuriy Makhlin, Gerd Sch\"{o}n, and Alexander
Shnirman, \textquotedblleft Quantum-state engineering with Josephson-junction
devices\textquotedblright, Rev. Mod. Phys. 73, 357(2001).

\bibitem {Inverse ac JosephsonE}M.T. Levinsen, R.Y. Chiao, M.J. Feldman, and
B.A. Tucker, \textquotedblleft An inverse AC Josephson effect voltage
standard,\textquotedblright\ Appl. Phys. Lett. \textbf{31}, 776 (1977).

\bibitem {QuantumCircuit}Bernard Yurke and John S. Denker, \textquotedblleft
Quantum network theory\textquotedblright\ ,Phys. Rev. A 29, 1419(1984).

\bibitem {Devoret}Michael H.Devoret, \textquotedblleft Quantum Fluctuations in
Electrical Circuits\textquotedblright, Quantum Fluctuations / Les Houches,
(Elsevier, Amsterdam, Netherlands, 1997) p. 351-86.

\bibitem {Flux qubit}J. E. Mooij, T. P. Orlando, L. Levitov, Lin Tian, Caspar
H. van der Wal, Seth Lloyd, \textquotedblleft Josephson Persistent-Current
Qubit\textquotedblright, Science 285, 1036 (1999).

\bibitem {Phase qubit}John M. Martinis, Michel H. Devoret, and John Clarke,
\textquotedblleft Energy-Level Quantization in the Zero-Voltage State of a
Current-Biased Josephson Junction\textquotedblright, Phys. Rev. Lett. 55, 1543(1985).

\bibitem {Hohensee}M.A. Hohensee, B. Estey, P. Hamilton, A. Zeilinger, and H.
M\"{u}ller, \textquotedblleft Force-free gravitational redshift: Proposed
gravitational Aharonov-Bohm experiment,\textquotedblright\ Phys. Rev. Lett.
108, 230404 (2012).

\bibitem {Tonomura}A. Tonomura, et al., \textquotedblleft Evidence for
Aharonov-Bohm Effect with Magnetic Field Completely Shielded from Electron
Wave,\textquotedblright\ Phys. Rev. Lett. 48\textbf{\ }(1982).

\bibitem {Kittel}C. Kittel,\emph{\ Introduction to Solid State Physics}. New
York: Wiley (1996). ISBN 0-471-14286-7.

\bibitem {Zeldovich}Ya. B. Zeldovich, \textquotedblleft The quasienergy of a
quantum mechanical system subjected to a periodic action,\textquotedblright J.
Exptl. Theoret. Phys. (U.S.S.R.) 51, 1492-1495 (November 1966); Soviet Physics
JETP 24, 1006 (1967).

\bibitem {Silveri}M. Silveri, J. Tuorila, M. Kemppainen, and E. Thuneberg,
\textquotedblleft Probe spectroscopy of quasienergy states,\textquotedblright%
\ arXiv:1301.0230 [quant-ph].

\bibitem {Wilczek}F. Wilczek, \textquotedblleft Quantum time
crystals,\textquotedblright\ Phys. Rev. Lett. 109, 160401 (2012).

\bibitem {Abramowitz}Abramowitz and Stegun (1972), p.361.
\end{thebibliography}
\end{document}